\documentclass[preprint,showpacs,showkeys,preprintnumbers,amsmath,amssymb]{revtex4}

\usepackage{graphicx,stmaryrd}
\usepackage{dcolumn}
\usepackage{bm}
\usepackage{textcomp} 
\usepackage{amsmath}
\usepackage{color}

\begin{document}

\newcommand{\C}{C$_{60}$ }

\preprint{JAP/Ambipolar OFETs}

\title{Ambipolar charge carrier transport in mixed organic
layers of phthalocyanine and fullerene}

\author{Andreas Opitz}
 \email[Corresponding author: ]{Andreas.Opitz@physik.uni-augsburg.de}

\author{Markus Bronner}%

\author{Wolfgang Br\"utting}

\affiliation{%
Experimentalphysik IV, Universit\"at Augsburg, 86135 Augsburg,
Germany
}%

\date{\today}

\begin{abstract}

Mixed layers of copper-phthalocyanine (p-conductive) and fullerene
(n-conductive) are used for the fabrication of organic
field-effect transistors (OFETs) and inverters. Regarding the
electrical characteristics of these donor-acceptor blends they
show ambipolar charge carrier transport, whereas devices made from
only one of the materials show unipolar behavior. Such mixed films
are model systems for ambipolar transport with adjustable
field-effect mobilities for electrons and holes. By variation of
the mixing ratio it is possible to balance the transport of both
charge-carrier types. In this paper we discuss the variation of
mobility and threshold voltage with the mixing ratio and
demonstrate ambipolar inverters as a leadoff application. The
gained results were analyzed by simulations using an analytical
model for ambipolar transistors and subsequently compared to
complementary inverters.

\end{abstract}



\maketitle

\section{Introduction}

In recent years organic semiconductors have attracted considerable
interest due to their growing potential as active materials in
electronic and optoelectronic devices. A long-standing paradigm,
however, has been seen in their unipolar transport of electrical
charges. This means that, apart from very few exceptions, organic
semiconductors have shown electrical conduction for one carrier
species only, with positive carriers being preferentially
transported in most materials. Thus multi-layer structures
comprising different organic materials with spatially separated
electron and hole transport layers where used for fabricating
efficient organic light-emitting diodes \cite{Tang87}. On the
other hand in photovoltaic devices, owing to short exciton
diffusion length in organic semiconductors, p- and n-conducting
materials need to be in close contact which is usually realized by
mixing them in one single layer yielding a so-called
bulk-heterojunction structure
\cite{meissner00,Yu95,Halls95,mix-stub,Uchida04}. Recently, such
donor-acceptor mixtures have been implemented also in organic
field-effect transistors (OFETs) and for the first time OFETs with
ambipolar characteristics were demonstrated \cite{Meij03}.
Meanwhile ambipolar OFETs have been realized with a variety of
material combinations, including polymer/fullerene blends
\cite{Meij03}, mixtures of soluble oligomers \cite{Lock03} as well
as evaporated molecular hetero-layer structures and mixed layers
\cite{Wang05,Xue04,Ye05}. However, even in neat films of a single
organic semiconductor ambipolar transport can be observed provided
that the barrier for electron injection is reduced by choosing
low-work function metals as source-drain electrodes \cite{TheoRS}
and that electron trapping at the interface to the gate-dielectric
is suppressed by surface functionalisation or suitable polymeric
dielectrics \cite{general-n}. Very recently, low-band gap
materials have been suggested as especially promising candidates
for ambipolar OFETs \cite{AmbiCirc}.

The successful fabrication of organic integrated circuits, which
are a key feature for realizing electronics based on organic
semiconductors, largely relies on the properties and functions of
OFETs as their basic constituents. So far, only unipolar logic
architectures consisting of p-conducting organic semiconductors
have evolved into a mature technology, mainly due to the
simplicity of circuit fabrication requiring only one
semiconducting layer and due fairly high environmental stability
of organic materials \cite{Gelinck04,Knobloch04}. However,
unipolar logic suffers from a number of drawbacks, such as low
performance, limited robustness and high power consumption. As in
silicon microelectronics, these problems in principle can be
overcome by using complementary logic circuits comprising p- and
n-channel transistors. Although, the proof-of-principle of this
approach has already been demonstrated \cite{Crone00}, the
fabrication of spatially separated organic p- and n-channel
transistors in an integrated circuit with micron-sized features
still appears to be a challenge. Thus, ambipolar OFETs which are
fabricated almost as simple as their unipolar counterparts might
be of interest for realizing complementary-like organic integrated
circuits \cite{AmbiCirc}. Moreover, it has been revealed that
ambipolar OFETs can provide additional functionalities as
light-emitting transistors \cite{AmbiOpto,Rost04} or
photo-transistors.

Therefore one intention of this paper is to compare ambipolar
inverters constructed of two ambipolar transistors with their
complementary counterparts consisting of two discrete p- and
n-channel transistors of the same materials. For this purpose we
prepare OFETs of pure and mixed semiconducting films made from
copper-phthalocyanine and buckminster-fullerene with various
mixing ratios and investigate their transport properties. The
results are related to the characteristics of ambipolar and
complimentary inverters and compared to simulations using an
analytical model. We will demonstrate that balanced mobilities of
electrons and holes can be achieved by varying the mixing ratio.
This turns out to be a crucial requirement for both ambipolar and
complementary logic devices.

\section{Device preparation and experimental methods}

Organic field-effect transistors were fabricated on highly
conductive Si wafers (1-5~m$\Omega$\,cm) with a 200~nm or 320~nm
thick thermally grown oxide, which acts as gate insulator.
Photolithographically patterned Ti(1~nm)/Au(100~nm) source and
drain electrodes were made by electron-beam evaporation and by
lift-off process. These structures were cleaned in an ultrasonic
bath with solvents (acetone, isopropyl) and ultra-pure water. The
substrates were dried with pure nitrogen, treated with an
O$_2$-plasma for 60~s at 200~W and 0.6~mbar, and heated in
fore-vacuum at 400~K for 2 hours.

\begin{figure}
\includegraphics[scale=0.5]{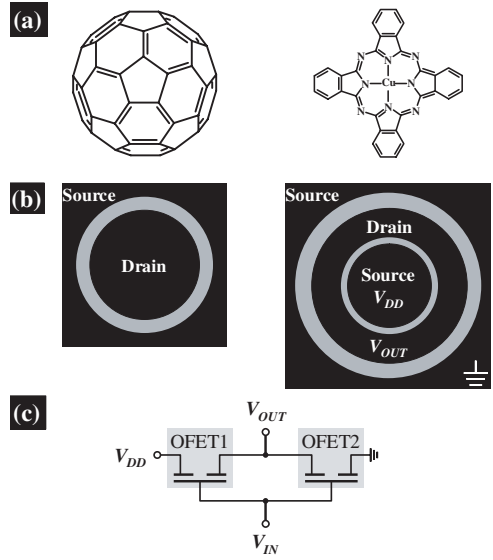} \caption{\label{fig:exp} (a) Chemical structures of the
used materials. Left: fullerene (C$_{60}$), right:
copper-phthalocyanine (CuPc). (b) Top view of the ring-type
transistor (left) and ring-type inverter (right). The silicon
substrate acts as the gate electrode in both cases. (c) Electrical
circuit of the used inverters.}
\end{figure}

Copper-phthalocyanine (CuPc, purchased from Aldrich as sublimation
grade) and buckminster fullerene (C$_{60}$, purchased from Hoechst
as super gold grade) were used as hole and electron conducting
materials, respectively. The structural formulas are given in
fig.~\ref{fig:exp}(a). The materials were deposited by thermal
evaporation from low-temperature effusion cells in a vacuum better
than $1\times 10^{-7}$~mbar to form pure and mixed layers with a
thickness of about 25~nm on the prestructured substrates. The
deposition rates were between 0.35~\AA/s for neat films and
1.4~\AA/s for layers with 1:3 stoichiometry. The given mixing
ratios in this paper are volume percentages as the evaporation
process was controlled via the deposition rates monitored with a
quartz microbalance. They are always given in the form
C$_{60}$:CuPc. Different substrate temperatures (300~K and 375~K)
were used during evaporation of the materials.

Transistors with a ring-type geometry (see fig.~\ref{fig:exp}(b),
left-hand side) were used, whose source electrodes form a closed
ring around the drain electrodes. This prevents parasitic currents
from outside of the active transistor channel without the
necessity of structuring the organic semiconductor
\cite{circ-design}. The channel length and width were 5~$\mu$m and
2500~$\mu$m, respectively. Ambipolar inverters were made from two
transistors stacked into each other (see fig.~\ref{fig:exp} (b)
and (c)) and have an additional ring channel around the first one.
Thus both transistors share the common silicon substrate gate
electrode as input of the inverter and the drain electrode as
output. Length and width of the outer channel were 10~$\mu$m and
2500~$\mu$m, respectively, and of the inner channel 8~$\mu$m and
2000~$\mu$m, respectively. Consequently the ratio of length to
width was the same for both channels. Each substrate (20~mm
$\times$ 20~mm) contained 24 individual transistors and 12
inverter structures to allow for a comparison of several devices.
Complementary inverters were fabricated by evaporating pure CuPc
and C$_{60}$ each on one half of a prestructured substrate. For
this purpose two evaporation steps were necessary with different
shadow masks. In this case two separated transistors, one p-type
and one n-type, from different areas on the substrate were
connected together to form an inverter.

Without air-exposure the devices were transferred to a
vacuum-chamber providing a pressure less than $5\times
10^{-6}$~mbar for characterization. The output and transfer
characteristics of the transistors were measured using two
independent source-meter units (Keithley 236). In order to measure
inverter transfer curves an additional source-meter unit (Keithley
2400) was implemented.

The charge-carrier mobilities $\mu$ and the threshold voltages
$V_{T}$ were extracted from the slope of the transfer
characteristics in the saturation region \mbox{$|V_D|>|V_G-V_T|$}
using the standard relationship:

\begin{equation}
  I_{D,sat} = \frac{W}{2L} \mu C_{Ox} \left( V_G-V_T \right)^2.
\end{equation}

Here $W$ is the channel width, $L$ is the channel length, $C_{Ox}$
is the gate-oxide capacitance per unit area, $V_G$ is the gate
voltage, and additionally $V_D$ is the drain voltage. Mobility
$\mu$ and threshold voltage $V_T$ were determined from the linear
regression of the measured data plotted as $\sqrt{I_{D,sat}}$ vs.
$V_G$. The inverter characteristics include the transfer curve
$V_{OUT}(V_{IN})$ as well as the current dissipation
$I_{DD}(V_{IN})$.

X-ray diffraction (XRD) patterns were obtained in $\theta-2
\theta$ geometry using a Siemens D-500 diffractometer (Cu
K$_\alpha$ radiation with a wavelength of 0.1542~nm) for analyzing
the crystallinity of the films. The morphology investigations were
performed using a Thermo Microscopes Auto Probe scanning force
microscope (SFM) operated in non-contact (tapping) mode.

\section{Results and discussion}

\subsection{Unipolar field-effect transistors}

\begin{figure*}
\includegraphics[scale=0.5]{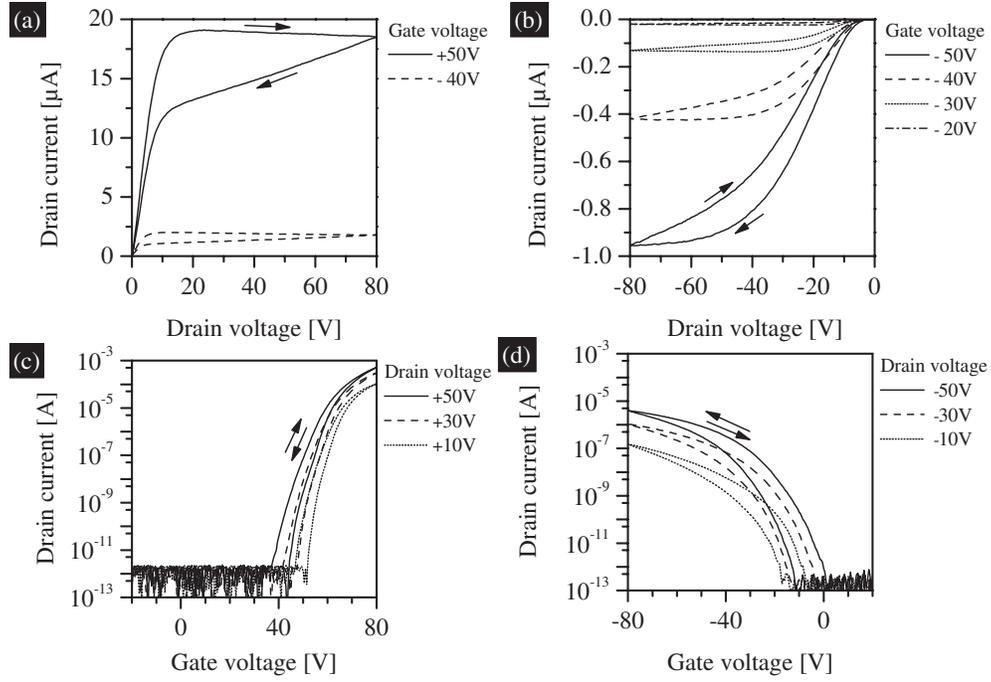} \caption{\label{fig:uni} Output characteristics of  unipolar
field-effect transistors with C$_{60}$ (a) and CuPc (b) and
corresponding transfer characteristic for C$_{60}$ (c) and for
CuPc (d). The substrates were O$_2$-plasma treated and the films
 evaporated at 375~K substrate temperature. The direction of the hysteresis is indicated
by arrows.}
\end{figure*}

Output and transfer characteristics of pure C$_{60}$ and CuPc FETs
are shown in fig.~\ref{fig:uni}. The output characteristics show
the typical unipolar transistor behavior with linear increase at
low drain voltage and saturation at higher $V_D$. The transfer
characteristics display an off-regime followed by an increase of
drain current with increasing absolute value of the gate voltage
exceeding the switch-on voltage.

The analysis of the transfer characteristics yielded saturation
mobilities for these preparation conditions (O$_2$ plasma
treatment, 375~K during evaporation) of
\mbox{$7\times10^{-2}$~cm$^2$/Vs} for the C$_{60}$ FET and
\mbox{$6\times10^{-4}$~cm$^2$/Vs} for the CuPc FET. The threshold
voltages were $+$63~V and $-$31~V.

The measured $I-V$ characteristics show considerable hysteresis
between increasing and decreasing voltage sweeps which can be
attributed to dynamic processes in the charging of the
semiconductor-dielectric interface \cite{Marjanovic06}, but in
this paper will not be discussed furthermore. The largest
hysteresis is present in the output characteristics of the \C FET.
The hysteresis in all other measurements, including the ambipolar
FETs, and in particular for the transfer characteristics which
were used for the analysis, are significantly smaller.

The curvature in the output characteristics of CuPc at the origin
of the $I$-$V$ diagram suggests an injection barrier with a
non-linear contact resistance whereas the linear increase in the
C$_{60}$ FET indicates a smaller injection barrier with a linear
contact resistance. This is related to the electronic structure of
the Au/CuPc and the Au/C$_{60}$ interface
\cite{E01,E02,E03,E04,GapCuPc,GapC60}. The injection barrier for
electrons from gold into C$_{60}$ is about 0.4~eV whereas the
value for holes into CuPc is significantly larger (about 0.9~eV).

\subsection{Ambipolar field-effect transistors}

\begin{figure*}
\includegraphics[scale=0.5]{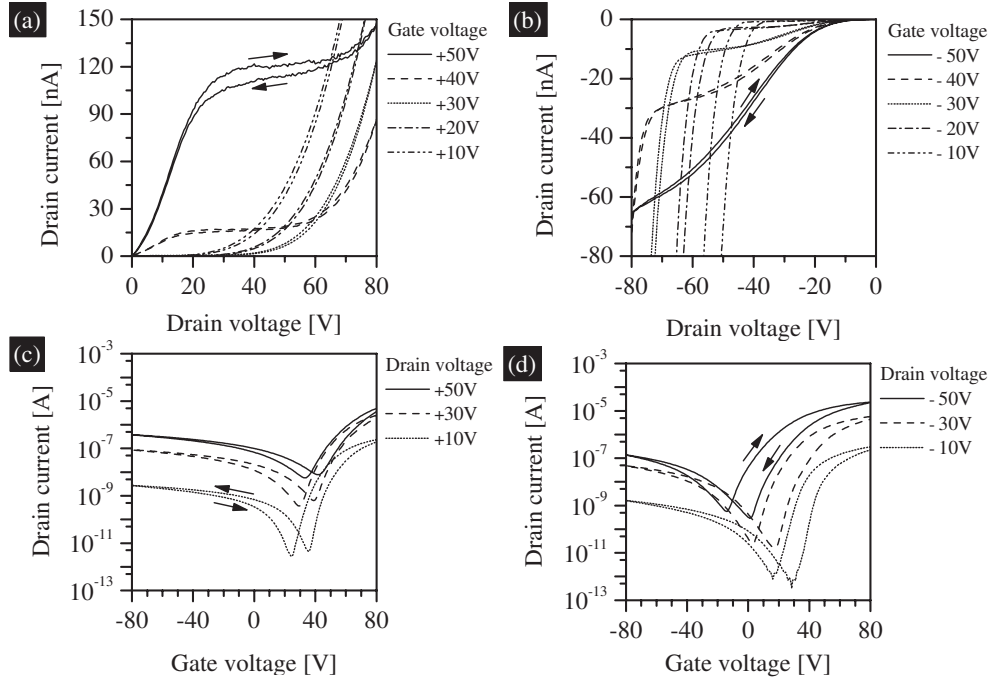}
\caption{\label{fig:ambi} Output characteristics of ambipolar
field-effect transistors for a mixing ratio between C$_{60}$ and
CuPc of 1:1 in the n-channel (a) and the p-channel regime (b)
together with the corresponding transfer characteristics in the
n-channel (c) and the p-channel regime (d). The substrate was
O$_2$-plasma treated and the film was evaporated at 375~K
substrate temperature. The direction of the hysteresis is
indicated by arrows.}
\end{figure*}

Ambipolar FETs with mixing ratios of 1:3, 1:1 and 3:1 between \C
and CuPc have been investigated. All of them showed ambipolar
transport with the same qualitative features as the 1:1 mixture
displayed in Figure~\ref{fig:ambi}. Here a strong increase of the
current in saturation of the output characteristic
(Figure~\ref{fig:ambi} a, b) is measured for both the n- and the
p-channel regime. This is a clear signature of ambipolar behavior.
Electrons are injected at higher drain voltages into the
hole-conducting channel, and vice versa. Consequently, the
transfer characteristics (Figure~\ref{fig:ambi} c, d) do not show
any off regime as the ambipolar increase takes place in the regime
where the FET with the pure materials is switched off.

The magnitude of the drain currents in  both regimes differs
significantly for this mixing ratio. The p-channel (negative
$V_G$) shows a current which is three orders of magnitude lower
than the current in the n-channel (positive $V_G$). In both
measurements the curvature in the output characteristics at the
origin of the $I$-$V$ diagram suggests injection barriers with
non-linear contact resistances. Here this contact resistance is
also visible in the electron channel and even more pronounced in
the hole channel because the conductivity of both channels is
reduced due to mutual dilution of the conducting material by the
other species. The nonlinear regime is then widened if the same
injection barrier is assumed.

\begin{figure*}
\includegraphics[scale=0.5]{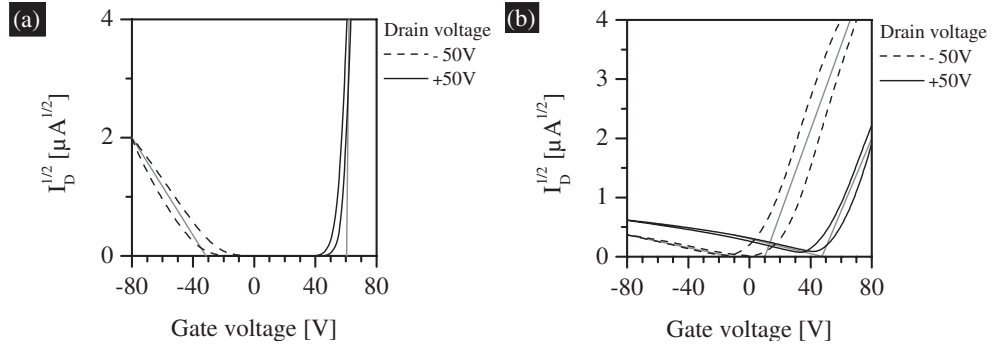}
\caption{\label{fig:trans} Square root of the drain current vs.
the gate voltage of unipolar field-effect transistors with
C$_{60}$ and CuPc (a) and the film with a mixing ratio of 1:1
(with raw data from fig.~\ref{fig:uni} and \ref{fig:ambi}). The
gray lines are linear fits from which the mobilities were
determined.}
\end{figure*}

Fig.~\ref{fig:trans} shows the analysis of the transfer
characteristics as the square root of the drain current vs. the
gate voltage of films with a mixing ratio of 1:1 as compared to
neat films of \C and CuPc . From the slope being directly
proportional to the square root of the mobility a decrease of the
electron and hole mobility is visible in the mixed film in
comparison to the mobilities in the pure films. This is a general
feature of mixed CuPc/\C layers and will be discussed more
detailed in the following subsection.

\subsection{\label{sec:mob}Dependence of charge carrier mobility on composition and thermal treatment}

\begin{figure}
\includegraphics[scale=0.5]{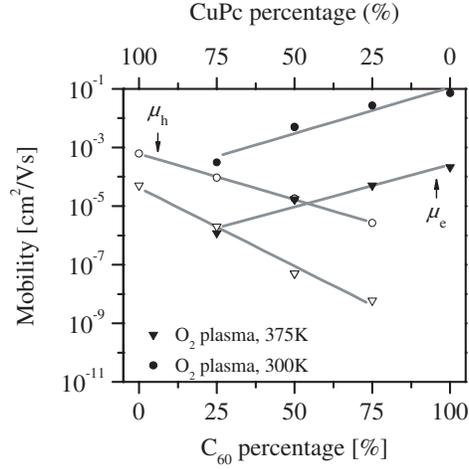}
\caption{\label{fig:mob} Dependence of the mobility on the mixing
ratio for different substrate temperatures during organic film
deposition (300~K and 375~K). The mobilities are determined from
the measurements in the saturation regime. Open symbols are
related to the hole mobilities and filled symbols to the electron
mobilities. The gray lines are linear fits conducing to
explicitness.}
\end{figure}

Figure~\ref{fig:mob} shows the variation of electron and hole
mobilities with the mixing ratio for two different substrate
temperatures during evaporation (300~K and 375~K). Apparently, an
exponential decrease of both $\mu_e$ and $\mu_h$ is observed upon
dilution of the corresponding conducting material with the other
species. A similar decrease has been reported in the literature
for an ambipolar light-emitting system of co-evaporated molecular
materials \cite{AmbiOpto}. Furthermore, we find that heating the
substrate during evaporation increases the mobilities of neat
films and mixed layers by up to three orders of magnitude.
Preliminary investigations have shown that a further increase is
obtained if substrate heating is combined with silanisation of the
SiO$_2$ surface prior to the deposition of the organic materials.
Mobilities of up to 1~cm$^2$/Vs and $2\times10^{-2}$~cm$^2$/Vs
were achieved for neat C$_{60}$ and CuPc films, respectively.
Remarkably, as the data in fig.~\ref{fig:mob} show, there is a
mixing ratio where the mobilities of electrons and holes are
equal. The films with balanced mobilities are CuPc-rich with about
75\% CuPc and 25\% C$_{60}$ content, which seems to be related to
the lower mobilities in neat CuPc films in comparison to C$_{60}$
films. We note that this mixing ratio for balanced mobilities of
about 3:1 is even valid for layers deposited on silanised SiO$_2$,
where a balanced mobility of $1\times10^{-2}$~cm$^2$~cm$^2$/Vs is
achieved in this case.

In order to get more insight into the concentration dependent
electron and hole mobility, structure and morphology of neat and
mixed films (evaporated at 375~K) were investigated by x-ray
diffraction and scanning force microscopy. XRD measurements (not
displayed here) of pure CuPc films show a strong (200) peak
corresponding to the $\alpha$-phase, whereas for all mixtures
including pure \C films no diffraction peaks are detectable,
indicating an amorphous structure. These observations are in full
agreement with measurements reported on CuPc:\C mixtures for
photovoltaic applications by Rand et al. \cite{XRD_Mix}. The
authors also observed a disappearance of the CuPc diffraction peak
for a fullerene content of more than about 15\% and asserted no
peak corresponding to C$_{60}$, although, like in our case, the
electron mobility in neat \C films was about two orders of
magnitude higher than the hole mobility in CuPc. Other groups have
obtained crystalline \C films, however, only at elevated substrate
temperatures of about 440~K during evaporation \cite{XRD_C60}.

\begin{figure*}
\includegraphics[scale=0.5]{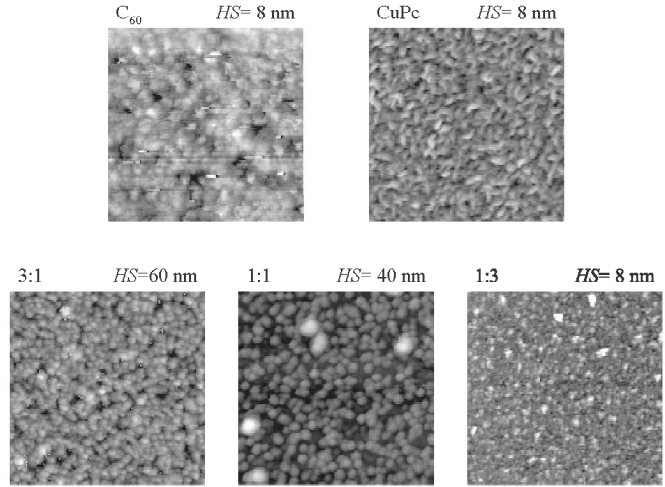}
\caption{\label{fig:sfm} Scanning force microscopy images taken in
non-contact mode for  pure C$_{60}$ and CuPc films as well as for
three mixed films grown at 375~K. The total image size is
\mbox{$1.9 \times 1.9~\mu\mathrm{m}^2$}. The height scale ($HS$)
is given as the difference between the lowest value (black) and
the highest value (white) in each of the images.}
\end{figure*}

Fig.~\ref{fig:sfm} shows the surface morphologies for different
mixing ratios as investigated by non-contact SFM. Neat CuPc films
have a needle-like structure, corresponding to the $\alpha$-phase
structure observed in XRD. For pure \C films a granular structure
is observed consisting of amorphous grains. Both films are
relatively smooth with a height scale of about 8~nm. This also
holds for CuPc-rich films with a mixing ratio of 1:3, however, the
grain size is now significantly reduced as compared to pure \C
films and smaller than the needles in pure CuPc films. It is to
note that this composition has balanced mobilities for electrons
and holes. The other two films with higher fullerene content
(mixing ratios of 1:1 and 3:1) are much rougher as the height
scale in the SFM images increases to 40~nm and 60~nm,
respectively. Consider that the nominal film thickness is only
about 25~nm. This observation can be related to demixing and phase
separation between \C and CuPc at elevated deposition temperature
\cite{demixing}.

At this point it is noteworthy to compare our results on the
variation of charge carrier mobilities with composition of the
blends to other ambipolar mixed systems. There are on the one hand
polymer-fullerene blends used for photovoltaic applications (in
particular MDMO-PPV:PCBM) \cite{Tuladhar05,Mihailetchi05} and on
the other hand blends of molecular materials implemented again in
photovoltaic devices (mostly phthalocyanine:\C) \cite{XRD_Mix} and
ambipolar light-emitting transistors \cite{Dinelli06}. In both
types of blend systems one observes as a common feature a strong
(in many cases exponential) decrease of the mobility of one
carrier type with increasing dilution by the other component. A
remarkable exception from this rule is the hole mobility in
polymer-fullerene blends in which the addition of fullerene
molecules even improves $\mu_h$. In this case a stretching of the
polymer chains is discoverable which can be put down to the
presence of PCBM leading to improved interchain hopping
\cite{Mihailetchi05}. There is not much of an interaction among
materials in low-molecular blends, they rather tend to dilute each
other. The fact that the average distance between molecules of the
same type in mixed donor-acceptor films is slightly increased
leads to a reduction of hopping probability and thereby compared
to neat films of one material only also to a reduction of the
charge carrier mobility. It is remarkable, however, that despite
of a roughness which is larger than the film thickness in some of
our films, all of them show n-channel as well as p-channel
conductivity. Obviously, there still exists a percolation path for
conduction in both materials independent of phase separation and
order formation. We also note that the bulk morphology does not
necessarily provide precise information for organic field-effect
transistors since in these devices the active channel is
restricted to the first few molecular layers at the interface to
the gate dielectric \cite{Dinelli04}. Nevertheless our results are
conform to the bulk mobility measurements by Rand et al.
\cite{XRD_Mix}.

\begin{figure}
\includegraphics[scale=0.5]{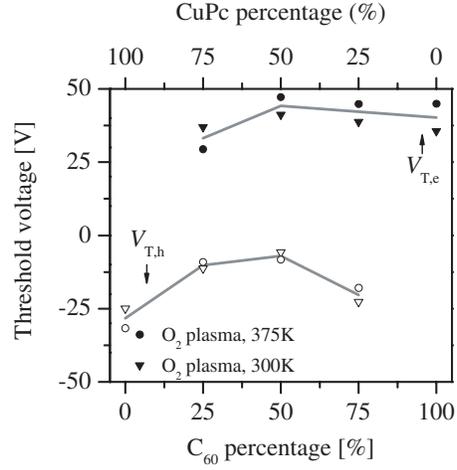}
\caption{\label{fig:vth} Dependence of the threshold voltage on
the mixing ratio for different temperature during organic film
evaporation (300~K and 375~K). The threshold voltages are
determined from the measurements in the saturation regime. Open
symbols are related to the threshold voltages for the holes and
filled symbols to the threshold voltages for the electrons. The
gray lines are connecting the average values for both treatments
and conduce to clarity.}
\end{figure}

Figure~\ref{fig:vth} shows the threshold voltages in dependence on
the mixing ratio and the substrate temperature (300~K and 375~K)
during evaporation of the organic films. For both regimes (n- and
p-channel), the threshold voltage has the most positive value at a
mixing ratio of 1:1. Also, there is no significant difference upon
thermal treatment. Apparently, the change of the threshold
voltages is stronger for the p-channel than for the n-channel
regime. A possible explanation can be found in the nature of
CuPc/C$_{60}$ blends being donor-acceptor systems. Measurements by
photoelectron spectroscopy \cite{E04} have shown that there is an
electron transfer from CuPc to C$_{60}$ leading to band bending
with hole accumulation at the CuPc side of the CuPc/C$_{60}$
interface, whereas no band bending arises at the C$_{60}$ side.
Although the reason for this has not yet been worked out in
detail, the hole accumulation in CuPc at the organic/organic
interface would explain a more positive threshold voltage for the
n-channel regime as can be seen in fig.~\ref{fig:vth}
\cite{Wang05}. Furthermore, it is plausible that the highest
charge transfer occurs at the mixing ratio of 1:1 where the
largest organic/organic interface area exists. The threshold
voltage shift will decrease in accordance with the interfacial
area for increasing fraction of one of the two materials. Because
the hole accumulation is on the CuPc side of the organic/organic
interface and there is no electron accumulation in C$_{60}$ the
change of the threshold voltage for the electron channel is
expected to be much weaker than for the hole channel.

The threshold voltage can be related to the work function
difference of metal and semiconductor, the doping concentration,
the interfacial charge density and the bulk potential
\cite{Delta_VT}. Considering only interface charges $N_{I\!f}$, as
all other parameters are not expected to change upon mixing of
molecules, the threshold voltage shift can be described by

\begin{equation}\label{eq:deltavt}
    \Delta V_T = \frac{e \cdot N_{I\!f}}{C_{Ox}}
\end{equation}

The observed threshold voltage shift of about 21~V at a mixing
ratio of C$_{60}$:CuPc of 1:1 results then in an interface charge
of 0.024 charges per square nanometer (calculated for a 200~nm
oxide). Corresponding to a charge transfer of 0.012 charges per
molecule. This calculation is based on a uniform material
distribution at the organic/insulator interface and a CuPc spacing
comparable to the $\alpha$-phase, with approximately two CuPc
molecules per square nanometer. The value mentioned is much less
than the reported charge transfer of 0.25 charges per molecule
determined from Raman spectroscopy \cite{Ruani02}, however, due to
the granular film morphology the calculated value can only be
regarded as a crude estimate. Further information concerning the
detailed molecular arrangement at the interface to the SiO$_2$
gate dielectric requires a specific quantitative analysis.

\subsection{\label{sec:model} Analytical model}

Theoretical studies of ambipolar transport in FETs were performed
for hydrogenated amorphous silicon (\textit{a}-Si:H)
\cite{TheoNeu1,TheoNeu2} where ambipolar transport was observed
for the first time \cite{a-Si-H}. Those models were useful for
describing measured characteristics but not for the extraction of
microscopic transport parameters. More recently, two-dimensional
simulations in organic system were performed by Paasch et al.
\cite{TheoGP}. Their simulations show that ambipolar transport is
possible for parameters allowing the injection of both charge
carrier types. An analytical model including microscopic transport
parameters was developed by Schmechel et al. \cite{TheoRS}. It is
an extension of the standard Shockley model for inorganic FETs and
allows for the determination of electron and hole mobilities and
the respective threshold voltages. As such, it assumes band
transport as in inorganic single-crystalline semiconductors and
neglects hopping transport as found in organic systems. The latter
problem recently has been treated by Anthopoulos et al. who
developed a consistent description for ambipolar OFETs based on
variable range hopping, including such important features like
density dependent charge carrier mobility \cite{Smits06}. However,
for the purpose of this paper the analytical model by Schmechel
turned out to be sufficient in order to work out the important
features of those different devices.

In this model the drain current of the electrons ($I_{D,e}$) and
of the holes ($I_{D,h}$) is given by \cite{TheoRS}:

\begin{widetext}
\begin{eqnarray}
I_{D,e} =  \frac{W}{2L} C_{Ox} \times \left\{ \mu_e \left(
\llbracket V_G - V_{T,e} \rrbracket^2 - \llbracket \left(V_G -
V_{T,e}\right) - V_D \rrbracket^2 \right) + \mu_h \left(
\llbracket V_D - \left(V_G -
V_{T,h} \right) \rrbracket^2 \right) \right\} \nonumber\\
I_{D,h} = -\frac{W}{2L} C_{Ox} \times \left\{ \mu_h \left(
\llbracket V_{T,h}-V_G \rrbracket^2 - \llbracket  V_D - \left( V_G
- V_{T,h} \right) \rrbracket^2 \right) + \mu_e \left( \llbracket
\left(V_G - V_{T,e} \right) - V_D\rrbracket^2 \right) \right\}
\label{eq:RSn}
\end{eqnarray}
\end{widetext}

Therein double brackets are defined as $\llbracket x
\rrbracket=\frac{1}{2}x-\frac{1}{2}\left|x\right|$ and allow for a
full analytical expression instead of a stepwise defined function.
The pre-factor includes the specific insulator capacitance
($C_{Ox}$) and the geometry of the FET ($W$ - channel width, $L$ -
channel length). The expression in curly brackets contains two
parts, one for the normal behavior of a unipolar FET (with $\mu_e$
as electron or $\mu_h$ as hole mobility) and another one for the
quadratic ambipolar increase including the mobility and the
threshold voltage of the opposite carrier type. $V_D$ and $V_G$
are the applied drain and gate voltage, respectively, and
$V_{T,e}$ and $V_{T,h}$ the threshold voltages for electrons and
holes, respectively. This model is based on homogeneous organic
layers and describes both unipolar and ambipolar behavior using
transport parameters of both carrier types. It implicitly assumes
complete recombination of positive and negative charge carriers.

The model has been used to fit output and transfer characteristics
of our samples. As an example, we have included in
fig.~\ref{fig:trans} the corresponding fit curves from which the
respective mobilities and threshold voltages have been derived
(see tab.~\ref{tab}). It is remarkable that the mobilities as
determined either from the saturation regime ($V_G > 0$, $V_D > 0$
for electrons and $V_G < 0$, $V_D < 0$ for holes, respectively)
and from the ambipolar regime ($V_G > 0$, $V_D < 0$ for electrons
and $V_G < 0$, $V_D > 0$ for holes, respectively) show excellent
agreement. Only for the threshold voltages there is a somewhat
larger difference between the two regimes. The origin of this
phenomenon is still being investigated. Nevertheless, this model
has proved to also be very useful for the simulation of ambipolar
and complementary inverters as described in the following.

\subsection{Measurements of ambipolar and complementary inverters}

As already mentioned, ambipolar transport is involved in various
organic semiconductor devices, including photovoltaic cells and
light-emitting FETs. Additionally there was the suggestion of
using ambipolar FETs to realize complementary-like organic
integrated circuits \cite{Meij03,AmbiCirc}. Here we investigate
ambipolar inverters consisting of mixed-layer ambipolar FETs and
compare their characteristics to a complementary inverter made of
discrete p- and n-channel transistors from neat materials.

\begin{figure*}
\includegraphics[scale=0.5]{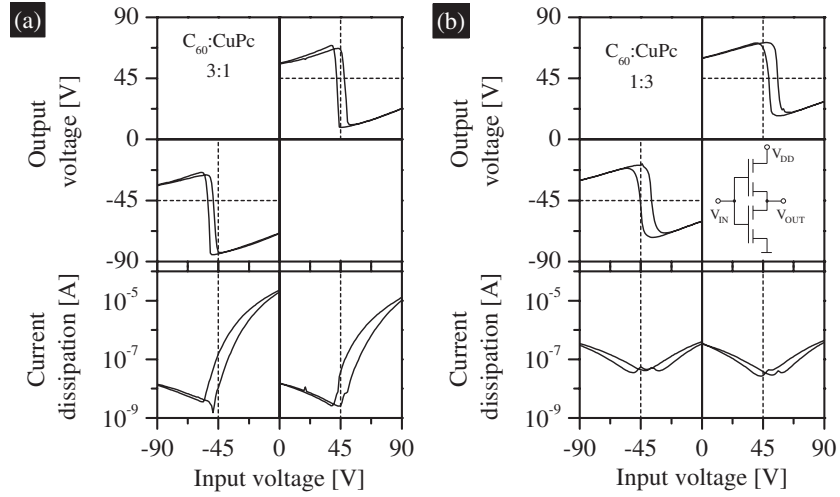}
\caption{\label{fig:ambi_inv}Characteristics of  ambipolar
inverters with  mixing ratios of 3:1 (a) and 1:3 (b) at a driving
voltage $V_{DD}=\pm$ 90~V. Transfer characteristics (top) and
current dissipation (bottom) are shown. The substrates were
O$_2$-plasma treated and the films were evaporated at 375~K
substrate temperature.}
\end{figure*}

Figure~\ref{fig:ambi_inv} shows the characteristics of ambipolar
inverters using layers with mixing ratios of 3:1 and 1:3. The
upper part presents the transfer characteristics of the inverters
(output voltage vs. input voltage) and the lower part the
corresponding current supplied by $V_{DD}$ which directly gives
the dissipated current in the circuit.

A characteristic feature of ambipolar inverters is their operation
in the first as well as in the third quadrant of the
output-vs.-input diagram, depending on the sign of the supply
voltage only. Ideally an inverter should have a sharp transition
from the low to the high state at half of the driving voltage and
the dissipated current should be negligibly small except for a
narrow range around the transition voltage. Both inverters, based
on two ambipolar transistors, show this transition at about half
of the supply voltage ($V_{DD}/2 = \pm 45$~V) and reach high gain
(about 13 for the 1:3 mixture and about 18 for the 3:1 mixture)
which is defined as the steepness of the characteristics at the
transitions between the high and the low states. However, they do
not reach zero in the low state and the driving voltage in the
high state. Also the voltages at the high and the low state are
not constant. The noise margin, which is an indicator for the
tolerance of cascaded inverter stages, is also high for both
inverters (about 14~V for the 1:3 mixture and about 19~V for the
3:1 mixture). Significant differences between the two ambipolar
inverters are observed in the dissipated current. Whereas for the
inverter with a 1:3 mixing ratio the power dissipation is
symmetric around $\pm$45~V, the device with a 3:1 mixture shows an
asymmetry of about three orders of magnitude. Moreover, the
dissipated current is as high as $10^{-5}$~A, which is
one-and-a-half orders of magnitude larger than in the previous
case. Thus, it is remarkable that a huge asymmetry in electron and
hole mobilities of more than three orders of magnitude, as
observed for the 3:1 mixture (see fig.~\ref{fig:mob}), has drastic
consequences for the power dissipation of the inverter, although
it does not lead to a significant asymmetry in the transition
voltage.

\begin{figure}
\includegraphics[scale=0.5]{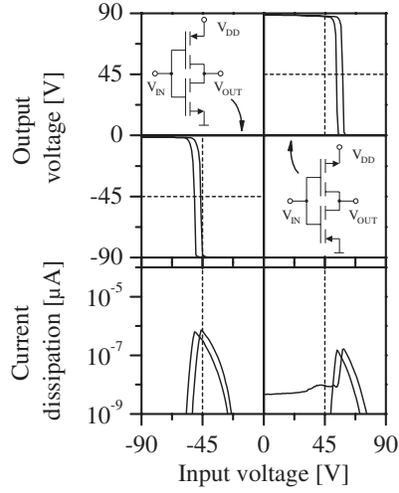}
\caption{\label{fig:comp_inv}Transfer characteristics (top) and
current dissipation (bottom) of a complementary inverter
consisting of discrete CuPc and \C transistors. The insets present
the circuit and the sign of the driving voltage used to operate
the inverter in the first and the third quadrant. The substrate
was O$_2$-plasma treated and the film was evaporated at 375~K
substrate temperature. }
\end{figure}

For comparison we have also fabricated a complementary inverter by
connecting a p-channel transistor (pure CuPc) and an n-channel
transistor (pure C$_{60}$) together, its characteristics being
shown in fig.~\ref{fig:comp_inv}. In order to make it operate in
the first quadrant it is necessary to connect the p-channel
transistor to $+V_{DD}$ and the n-channel transistor to ground and
vice versa the n-channel transistor to $-V_{DD}$ and the p-channel
transistor to ground in order make it operate in the third
quadrant. These inverters also show slightly asymmetric
transitions with respect to $\pm V_{DD}/2$ due to the unbalanced
electron and hole mobilities in pure CuPc and C$_{60}$, but they
reach the ground potential in the low state and the supply voltage
in the high state. The gain is about 38 and the noise margin
between 30 and 50~V. A characteristic difference to the ambipolar
inverter type is the current dissipation being high only in the
vicinity of the transition. This is because at any time one of the
two transistors is being switched off in each of the logic states
whereas an ambipolar transistor always shows a non-negligible
current. We note that the constant current of about $10^{-8}$~A
which is detected for positive driving voltage between 0 and about
45~V is not an inverter failure. We believe that this feature is
related to the large hysteresis of the C$_{60}$ FET (see fig.
\ref{fig:uni}), because it is visible also for other driving
voltages and occurs only during decreasing input voltage sweeps.

\subsection{\label{sec:sim}Simulation of ambipolar and complementary inverters}

For a better understanding of inverter characteristics we
performed numerical simulations based on the analytical model of
Schmechel et al. \cite{TheoRS} (see eq.~\ref{eq:RSn}) for both
ambipolar and complementary inverters. The output voltage as well
as the current dissipation was calculated to demonstrate the
differences between ambipolar and complementary inverters and the
influence of device parameters like mobility and threshold
voltage.

\begin{figure*}
\includegraphics[scale=0.5]{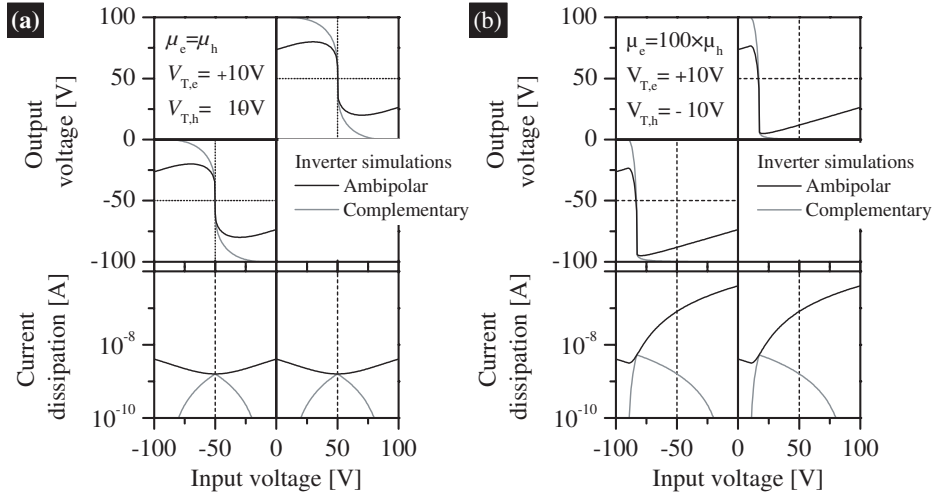}
\caption{\label{fig:sim_inv}Simulations of ambipolar and
complementary inverter transfer characteristics and current
dissipation with symmetric parameters (mobility and threshold
voltage) for the p- and the n-channel in part (a), in comparison
to asymmetric mobilities and symmetric threshold voltages in part
(b).}
\end{figure*}

Figure~\ref{fig:sim_inv} shows simulated transfer characteristics
of ambipolar and complementary inverters for balanced and
unbalanced charge carrier mobilities. In both cases, as observed
before in the measurements, the output voltage of the
complementary inverter reaches the driving and the ground voltage
in the high and the low state, respectively, whereas the output
voltage of the ambipolar inverter does not. Thus the noise margin
of the ambipolar inverter is lower than for the complementary
inverter. Nevertheless the gain for both types of inverters is
comparably high. As expected the current dissipation in the
complementary inverter is highest at the transition between the
logic levels, whereas the ambipolar inverter has its minimum at
this point.

A closer look at the simulations in fig.~\ref{fig:sim_inv}(a),
which are based on symmetric mobilities and threshold voltages,
shows that the transition between the logic states is at the half
of the driving voltage for both the ambipolar and the
complementary inverter. An asymmetry in carrier mobilities
(fig.~\ref{fig:sim_inv}(b)) and/or in threshold voltages (not
shown here) leads to a shift of the transition voltage in both
cases and a higher current dissipation, particularly for the
ambipolar case. This emphasizes the need for balanced electron and
hole mobilities in both inverter types.

\section{Summary}

Ambipolar transport in mixed films of C$_{60}$ and CuPc was
investigated in field-effect transistors and compared to unipolar
transport in neat films. Hole and electron transport is observed
for all analyzed mixing ratios and deposition temperature with
carrier mobilities decreasing exponentially for decreasing volume
percentage of the respective conducting material. A mixing ratio
of \C to CuPc of 1:3 is found to yield balanced electron and hole
mobilities for all preparation conditions. The variation of the
threshold voltage is tentatively explained by a hole accumulation
at the CuPc side of the organic/organic interface.

Furthermore we have demonstrated ambipolar inverters with
complementary-like logic behavior from such mixed layers and
compared their performance to complementary inverters consisting
of discrete p- and n-channel transistors of the neat materials.
Both our experimental results and the simulations using an
analytical model underline the need for balanced carrier
mobilities to achieve symmetric inverter characteristics and low
losses. While in a complementary inverter this has to be
accomplished by a proper choice of p- and n-conducting materials,
the approach using mixed layers, as demonstrated here, allows to
achieve this balance by adjusting the composition of the mixed
ambipolar layer.

\section*{Acknowledgements}

This work was supported by the Deutsche Forschungsgemeinschaft
(Focus Programme 1121 "Organic Field-Effect Transistors"). The
discussed x-ray diffraction measurements were kindly performed by
Jens Pflaum (University of Stuttgart).

\newpage

\begin{table}
  \begin{ruledtabular}
  \caption{Extracted parameter values (mobility and threshold voltage)
  for different mixing ratios determined from saturation and the ambipolar regime.
  Part of the fit curves is also shown in fig.~\ref{fig:trans}.}\label{tab}
  \begin{tabular}{|ccccc|}
                      &\mbox{$\mu_{e}$}&$\mu_{h}$&$V_{T,e}$&$V_{T,h}$\\
                      &\mbox{[cm$^2$/Vs]}&[cm$^2$/Vs]&$[V]$&$[V]$ \\\hline
\multicolumn{5}{|c|}{Determined from saturation regime}\\\hline
    pure C$_{60}$    &$3.2\times 10^{-1}$& ---               & $+60.4$ &  ---  \\
    C$_{60}$:CuPc 3:1&$2.2\times 10^{-2}$&$2.0\times 10^{-6}$& $+44.8$ & $-22.7$ \\
    C$_{60}$:CuPc 1:1&$1.3\times 10^{-3}$&$7.7\times 10^{-6}$& $+47.4$ & $-11.9$ \\
    C$_{60}$:CuPc 1:3&$3.5\times 10^{-4}$&$1.0\times 10^{-4}$& $+30.7$ & $-17.0$ \\
    pure CuPc        &         ---       &$8.5\times 10^{-3}$&  ---    & $-31.5$ \\\hline
\multicolumn{5}{|c|}{Determined from ambipolar regime}\\\hline
    C$_{60}$:CuPc 3:1&$2.7\times 10^{-2}$&$1.6\times 10^{-6}$& $+49.5$ & $-24.0$ \\
    C$_{60}$:CuPc 1:1&$1.6\times 10^{-3}$&$8.6\times 10^{-6}$& $+64.5$ & \hspace{1.5mm}$-1.7$ \\
    C$_{60}$:CuPc 1:3&$3.1\times 10^{-4}$&$1.0\times 10^{-4}$& $+35.1$ & $-17.7$ \\
  \end{tabular}
  \end{ruledtabular}
\end{table}

\clearpage

\clearpage

\clearpage

\clearpage

\clearpage

\clearpage

\clearpage

\clearpage

\clearpage

\clearpage

\begin{thebibliography}{41}
\expandafter\ifx\csname
natexlab\endcsname\relax\def\natexlab#1{#1}\fi
\expandafter\ifx\csname bibnamefont\endcsname\relax
  \def\bibnamefont#1{#1}\fi
\expandafter\ifx\csname bibfnamefont\endcsname\relax
  \def\bibfnamefont#1{#1}\fi
\expandafter\ifx\csname citenamefont\endcsname\relax
  \def\citenamefont#1{#1}\fi
\expandafter\ifx\csname url\endcsname\relax
  \def\url#1{\texttt{#1}}\fi
\expandafter\ifx\csname
urlprefix\endcsname\relax\def\urlprefix{URL }\fi
\providecommand{\bibinfo}[2]{#2}
\providecommand{\eprint}[2][]{\url{#2}}

\bibitem[{\citenamefont{Tang and Vanslyke}(1987)}]{Tang87}
\bibinfo{author}{\bibfnamefont{C.~W.} \bibnamefont{Tang}} \bibnamefont{and}
  \bibinfo{author}{\bibfnamefont{S.~A.} \bibnamefont{Vanslyke}},
  \bibinfo{journal}{Appl. Phys. Lett.} \textbf{\bibinfo{volume}{51}},
  \bibinfo{pages}{913} (\bibinfo{year}{1987}).

\bibitem[{\citenamefont{St{\"u}binger and Br{\"u}tting}(2002)}]{mix-stub}
\bibinfo{author}{\bibfnamefont{T.}~\bibnamefont{St{\"u}binger}}
  \bibnamefont{and}
  \bibinfo{author}{\bibfnamefont{W.}~\bibnamefont{Br{\"u}tting}},
  \bibinfo{journal}{SPIE Proc.} \textbf{\bibinfo{volume}{4465}},
  \bibinfo{pages}{102} (\bibinfo{year}{2002}).

\bibitem[{\citenamefont{Yu et~al.}(1995)\citenamefont{Yu, Gao, Hummelen, Wudl,
  and Heeger}}]{Yu95}
\bibinfo{author}{\bibfnamefont{G.}~\bibnamefont{Yu}},
  \bibinfo{author}{\bibfnamefont{J.}~\bibnamefont{Gao}},
  \bibinfo{author}{\bibfnamefont{J.~C.} \bibnamefont{Hummelen}},
  \bibinfo{author}{\bibfnamefont{F.}~\bibnamefont{Wudl}}, \bibnamefont{and}
  \bibinfo{author}{\bibfnamefont{A.~J.} \bibnamefont{Heeger}},
  \bibinfo{journal}{Science} \textbf{\bibinfo{volume}{270}},
  \bibinfo{pages}{1789} (\bibinfo{year}{1995}).

\bibitem[{\citenamefont{Halls et~al.}(1995)\citenamefont{Halls, Walsh,
  Greenham, Marseglia, Friend, Moratti, and Holmes}}]{Halls95}
\bibinfo{author}{\bibfnamefont{J.~J.~M.} \bibnamefont{Halls}},
  \bibinfo{author}{\bibfnamefont{C.~A.} \bibnamefont{Walsh}},
  \bibinfo{author}{\bibfnamefont{N.~C.} \bibnamefont{Greenham}},
  \bibinfo{author}{\bibfnamefont{E.~A.} \bibnamefont{Marseglia}},
  \bibinfo{author}{\bibfnamefont{R.~H.} \bibnamefont{Friend}},
  \bibinfo{author}{\bibfnamefont{S.~C.} \bibnamefont{Moratti}},
  \bibnamefont{and} \bibinfo{author}{\bibfnamefont{A.~B.}
  \bibnamefont{Holmes}}, \bibinfo{journal}{Nature}
  \textbf{\bibinfo{volume}{376}}, \bibinfo{pages}{498} (\bibinfo{year}{1995}).

\bibitem[{\citenamefont{Uchida et~al.}(2004)\citenamefont{Uchida, Xue, Rand,
  and Forrest}}]{Uchida04}
\bibinfo{author}{\bibfnamefont{S.}~\bibnamefont{Uchida}},
  \bibinfo{author}{\bibfnamefont{J.~G.} \bibnamefont{Xue}},
  \bibinfo{author}{\bibfnamefont{B.~P.} \bibnamefont{Rand}}, \bibnamefont{and}
  \bibinfo{author}{\bibfnamefont{S.~R.} \bibnamefont{Forrest}},
  \bibinfo{journal}{Appl. Phys. Lett.} \textbf{\bibinfo{volume}{84}},
  \bibinfo{pages}{4218} (\bibinfo{year}{2004}).

\bibitem[{\citenamefont{Rostalski and Meissner}(2000)}]{meissner00}
\bibinfo{author}{\bibfnamefont{J.}~\bibnamefont{Rostalski}} \bibnamefont{and}
  \bibinfo{author}{\bibfnamefont{D.}~\bibnamefont{Meissner}},
  \bibinfo{journal}{Sol. Energ. Mat. Sol. C.} \textbf{\bibinfo{volume}{61}},
  \bibinfo{pages}{87} (\bibinfo{year}{2000}).

\bibitem[{\citenamefont{Meijer et~al.}(2003{\natexlab{a}})\citenamefont{Meijer,
  De~Leeuw, Setayesh, Van~Veenendaal, Huisman, Blom, Hummelen, Scherf, and
  Klapwijk}}]{Meij03}
\bibinfo{author}{\bibfnamefont{E.~J.} \bibnamefont{Meijer}},
  \bibinfo{author}{\bibfnamefont{D.~M.} \bibnamefont{De~Leeuw}},
  \bibinfo{author}{\bibfnamefont{S.}~\bibnamefont{Setayesh}},
  \bibinfo{author}{\bibfnamefont{E.}~\bibnamefont{Van~Veenendaal}},
  \bibinfo{author}{\bibfnamefont{B.~H.} \bibnamefont{Huisman}},
  \bibinfo{author}{\bibfnamefont{P.~W.~M.} \bibnamefont{Blom}},
  \bibinfo{author}{\bibfnamefont{J.~C.} \bibnamefont{Hummelen}},
  \bibinfo{author}{\bibfnamefont{U.}~\bibnamefont{Scherf}}, \bibnamefont{and}
  \bibinfo{author}{\bibfnamefont{T.~M.} \bibnamefont{Klapwijk}},
  \bibinfo{journal}{Nature Mater.} \textbf{\bibinfo{volume}{2}},
  \bibinfo{pages}{678} (\bibinfo{year}{2003}{\natexlab{a}}).

\bibitem[{\citenamefont{Locklin et~al.}(2003)\citenamefont{Locklin, Shinbo,
  Onishi, Kaneko, Bao, and Advincula}}]{Lock03}
\bibinfo{author}{\bibfnamefont{J.}~\bibnamefont{Locklin}},
  \bibinfo{author}{\bibfnamefont{K.}~\bibnamefont{Shinbo}},
  \bibinfo{author}{\bibfnamefont{K.}~\bibnamefont{Onishi}},
  \bibinfo{author}{\bibfnamefont{F.}~\bibnamefont{Kaneko}},
  \bibinfo{author}{\bibfnamefont{Z.~N.} \bibnamefont{Bao}}, \bibnamefont{and}
  \bibinfo{author}{\bibfnamefont{R.~C.} \bibnamefont{Advincula}},
  \bibinfo{journal}{Chem. Mater.} \textbf{\bibinfo{volume}{15}},
  \bibinfo{pages}{1404} (\bibinfo{year}{2003}).

\bibitem[{\citenamefont{Wang et~al.}(2005)\citenamefont{Wang, Wang, Yan, Huang,
  and Yan}}]{Wang05}
\bibinfo{author}{\bibfnamefont{J.}~\bibnamefont{Wang}},
  \bibinfo{author}{\bibfnamefont{H.~B.} \bibnamefont{Wang}},
  \bibinfo{author}{\bibfnamefont{X.~J.} \bibnamefont{Yan}},
  \bibinfo{author}{\bibfnamefont{H.~C.} \bibnamefont{Huang}}, \bibnamefont{and}
  \bibinfo{author}{\bibfnamefont{D.~H.} \bibnamefont{Yan}},
  \bibinfo{journal}{Appl. Phys. Lett.} \textbf{\bibinfo{volume}{87}},
  \bibinfo{pages}{093507} (\bibinfo{year}{2005}).

\bibitem[{\citenamefont{Ye et~al.}(2005)\citenamefont{Ye, Baba, Oishi, Mori,
  and Suzuki}}]{Ye05}
\bibinfo{author}{\bibfnamefont{R.~B.} \bibnamefont{Ye}},
  \bibinfo{author}{\bibfnamefont{M.}~\bibnamefont{Baba}},
  \bibinfo{author}{\bibfnamefont{Y.}~\bibnamefont{Oishi}},
  \bibinfo{author}{\bibfnamefont{K.}~\bibnamefont{Mori}}, \bibnamefont{and}
  \bibinfo{author}{\bibfnamefont{K.}~\bibnamefont{Suzuki}},
  \bibinfo{journal}{Appl. Phys. Lett.} \textbf{\bibinfo{volume}{86}},
  \bibinfo{pages}{253505} (\bibinfo{year}{2005}).

\bibitem[{\citenamefont{Xue and Forrest}(2004)}]{Xue04}
\bibinfo{author}{\bibfnamefont{J.~G.} \bibnamefont{Xue}} \bibnamefont{and}
  \bibinfo{author}{\bibfnamefont{S.~R.} \bibnamefont{Forrest}},
  \bibinfo{journal}{Phys. Rev. B} \textbf{\bibinfo{volume}{69}},
  \bibinfo{pages}{245322} (\bibinfo{year}{2004}).

\bibitem[{\citenamefont{Schmechel et~al.}(2005)\citenamefont{Schmechel, Ahles,
  and von Seggern}}]{TheoRS}
\bibinfo{author}{\bibfnamefont{R.}~\bibnamefont{Schmechel}},
  \bibinfo{author}{\bibfnamefont{M.}~\bibnamefont{Ahles}}, \bibnamefont{and}
  \bibinfo{author}{\bibfnamefont{H.}~\bibnamefont{von Seggern}},
  \bibinfo{journal}{J. Appl. Phys.} \textbf{\bibinfo{volume}{98}},
  \bibinfo{pages}{084511} (\bibinfo{year}{2005}).

\bibitem[{\citenamefont{Chua et~al.}(2005)\citenamefont{Chua, Zaumseil, Chang,
  Ou, Ho, Sirringhaus, and Friend}}]{general-n}
\bibinfo{author}{\bibfnamefont{L.~L.} \bibnamefont{Chua}},
  \bibinfo{author}{\bibfnamefont{J.}~\bibnamefont{Zaumseil}},
  \bibinfo{author}{\bibfnamefont{J.~F.} \bibnamefont{Chang}},
  \bibinfo{author}{\bibfnamefont{E.~C.~W.} \bibnamefont{Ou}},
  \bibinfo{author}{\bibfnamefont{P.~K.~H.} \bibnamefont{Ho}},
  \bibinfo{author}{\bibfnamefont{H.}~\bibnamefont{Sirringhaus}},
  \bibnamefont{and} \bibinfo{author}{\bibfnamefont{R.~H.}
  \bibnamefont{Friend}}, \bibinfo{journal}{Nature}
  \textbf{\bibinfo{volume}{434}}, \bibinfo{pages}{194} (\bibinfo{year}{2005}).

\bibitem[{\citenamefont{Anthopoulos et~al.}(2006)\citenamefont{Anthopoulos,
  Setayesh, Smith, C\"{o}lle, Cantatore, de~Boer, Blom, and
  de~Leeuw}}]{AmbiCirc}
\bibinfo{author}{\bibfnamefont{T.~D.} \bibnamefont{Anthopoulos}},
  \bibinfo{author}{\bibfnamefont{S.}~\bibnamefont{Setayesh}},
  \bibinfo{author}{\bibfnamefont{E.}~\bibnamefont{Smith}},
  \bibinfo{author}{\bibfnamefont{M.}~\bibnamefont{C\"{o}lle}},
  \bibinfo{author}{\bibfnamefont{E.}~\bibnamefont{Cantatore}},
  \bibinfo{author}{\bibfnamefont{B.}~\bibnamefont{de~Boer}},
  \bibinfo{author}{\bibfnamefont{P.~W.~M.} \bibnamefont{Blom}},
  \bibnamefont{and} \bibinfo{author}{\bibfnamefont{D.~M.}
  \bibnamefont{de~Leeuw}}, \bibinfo{journal}{Adv. Mater.}
  \textbf{\bibinfo{volume}{18}}, \bibinfo{pages}{1900} (\bibinfo{year}{2006}).

\bibitem[{\citenamefont{Gelinck et~al.}(2004)\citenamefont{Gelinck, Huitema,
  Van~Veenendaal, Cantatore, Schrijnemakers, Van~der Putten, Geuns,
  Beenhakkers, Giesbers, Huisman et~al.}}]{Gelinck04}
\bibinfo{author}{\bibfnamefont{G.~H.} \bibnamefont{Gelinck}},
  \bibinfo{author}{\bibfnamefont{H.~E.~A.} \bibnamefont{Huitema}},
  \bibinfo{author}{\bibfnamefont{E.}~\bibnamefont{Van~Veenendaal}},
  \bibinfo{author}{\bibfnamefont{E.}~\bibnamefont{Cantatore}},
  \bibinfo{author}{\bibfnamefont{L.}~\bibnamefont{Schrijnemakers}},
  \bibinfo{author}{\bibfnamefont{J.}~\bibnamefont{Van~der Putten}},
  \bibinfo{author}{\bibfnamefont{T.~C.~T.} \bibnamefont{Geuns}},
  \bibinfo{author}{\bibfnamefont{M.}~\bibnamefont{Beenhakkers}},
  \bibinfo{author}{\bibfnamefont{J.~B.} \bibnamefont{Giesbers}},
  \bibinfo{author}{\bibfnamefont{B.~H.} \bibnamefont{Huisman}},
  \bibnamefont{et~al.}, \bibinfo{journal}{Nature Mater.}
  \textbf{\bibinfo{volume}{3}}, \bibinfo{pages}{106} (\bibinfo{year}{2004}).

\bibitem[{\citenamefont{Knobloch et~al.}(2004)\citenamefont{Knobloch, Manuelli,
  Bernds, and Clemens}}]{Knobloch04}
\bibinfo{author}{\bibfnamefont{A.}~\bibnamefont{Knobloch}},
  \bibinfo{author}{\bibfnamefont{A.}~\bibnamefont{Manuelli}},
  \bibinfo{author}{\bibfnamefont{A.}~\bibnamefont{Bernds}}, \bibnamefont{and}
  \bibinfo{author}{\bibfnamefont{W.}~\bibnamefont{Clemens}},
  \bibinfo{journal}{J. Appl. Phys.} \textbf{\bibinfo{volume}{96}},
  \bibinfo{pages}{2286} (\bibinfo{year}{2004}).

\bibitem[{\citenamefont{Crone et~al.}(2000)\citenamefont{Crone, Dodabalapur,
  Lin, Filas, Bao, LaDuca, Sarpeshkar, Katz, and Li}}]{Crone00}
\bibinfo{author}{\bibfnamefont{B.}~\bibnamefont{Crone}},
  \bibinfo{author}{\bibfnamefont{A.}~\bibnamefont{Dodabalapur}},
  \bibinfo{author}{\bibfnamefont{Y.~Y.} \bibnamefont{Lin}},
  \bibinfo{author}{\bibfnamefont{R.~W.} \bibnamefont{Filas}},
  \bibinfo{author}{\bibfnamefont{Z.}~\bibnamefont{Bao}},
  \bibinfo{author}{\bibfnamefont{A.}~\bibnamefont{LaDuca}},
  \bibinfo{author}{\bibfnamefont{R.}~\bibnamefont{Sarpeshkar}},
  \bibinfo{author}{\bibfnamefont{H.~E.} \bibnamefont{Katz}}, \bibnamefont{and}
  \bibinfo{author}{\bibfnamefont{W.}~\bibnamefont{Li}},
  \bibinfo{journal}{Nature} \textbf{\bibinfo{volume}{403}},
  \bibinfo{pages}{521} (\bibinfo{year}{2000}).

\bibitem[{\citenamefont{Rost et~al.}(2004)\citenamefont{Rost, Karg, Riess, Loi,
  Murgia, and Muccini}}]{Rost04}
\bibinfo{author}{\bibfnamefont{C.}~\bibnamefont{Rost}},
  \bibinfo{author}{\bibfnamefont{S.}~\bibnamefont{Karg}},
  \bibinfo{author}{\bibfnamefont{W.}~\bibnamefont{Riess}},
  \bibinfo{author}{\bibfnamefont{M.~A.} \bibnamefont{Loi}},
  \bibinfo{author}{\bibfnamefont{M.}~\bibnamefont{Murgia}}, \bibnamefont{and}
  \bibinfo{author}{\bibfnamefont{M.}~\bibnamefont{Muccini}},
  \bibinfo{journal}{Appl. Phys. Lett.} \textbf{\bibinfo{volume}{85}},
  \bibinfo{pages}{1613} (\bibinfo{year}{2004}).

\bibitem[{\citenamefont{Loi et~al.}(2006)\citenamefont{Loi, Rost-Bietsch,
  Murgia, Karg, Riess, and Muccini}}]{AmbiOpto}
\bibinfo{author}{\bibfnamefont{M.~A.} \bibnamefont{Loi}},
  \bibinfo{author}{\bibfnamefont{C.}~\bibnamefont{Rost-Bietsch}},
  \bibinfo{author}{\bibfnamefont{M.}~\bibnamefont{Murgia}},
  \bibinfo{author}{\bibfnamefont{S.}~\bibnamefont{Karg}},
  \bibinfo{author}{\bibfnamefont{W.}~\bibnamefont{Riess}}, \bibnamefont{and}
  \bibinfo{author}{\bibfnamefont{M.}~\bibnamefont{Muccini}},
  \bibinfo{journal}{Adv. Func. Mater.} \textbf{\bibinfo{volume}{16}},
  \bibinfo{pages}{41} (\bibinfo{year}{2006}).

\bibitem[{\citenamefont{Meijer et~al.}(2003{\natexlab{b}})\citenamefont{Meijer,
  Detcheverry, Baesjou, van Veenendaal, de~Leeuw, and Klapwijk}}]{circ-design}
\bibinfo{author}{\bibfnamefont{E.~J.} \bibnamefont{Meijer}},
  \bibinfo{author}{\bibfnamefont{C.}~\bibnamefont{Detcheverry}},
  \bibinfo{author}{\bibfnamefont{P.~J.} \bibnamefont{Baesjou}},
  \bibinfo{author}{\bibfnamefont{E.}~\bibnamefont{van Veenendaal}},
  \bibinfo{author}{\bibfnamefont{D.~M.} \bibnamefont{de~Leeuw}},
  \bibnamefont{and} \bibinfo{author}{\bibfnamefont{T.~M.}
  \bibnamefont{Klapwijk}}, \bibinfo{journal}{J. Appl. Phys.}
  \textbf{\bibinfo{volume}{93}}, \bibinfo{pages}{4831}
  (\bibinfo{year}{2003}{\natexlab{b}}).

\bibitem[{\citenamefont{Marjanovic et~al.}(2006)\citenamefont{Marjanovic,
  Singh, Dennler, Gunes, Neugebauer, Sariciftci, Schwodiauer, and
  Bauer}}]{Marjanovic06}
\bibinfo{author}{\bibfnamefont{N.}~\bibnamefont{Marjanovic}},
  \bibinfo{author}{\bibfnamefont{T.~B.} \bibnamefont{Singh}},
  \bibinfo{author}{\bibfnamefont{G.}~\bibnamefont{Dennler}},
  \bibinfo{author}{\bibfnamefont{S.}~\bibnamefont{Gunes}},
  \bibinfo{author}{\bibfnamefont{H.}~\bibnamefont{Neugebauer}},
  \bibinfo{author}{\bibfnamefont{N.~S.} \bibnamefont{Sariciftci}},
  \bibinfo{author}{\bibfnamefont{R.}~\bibnamefont{Schwodiauer}},
  \bibnamefont{and} \bibinfo{author}{\bibfnamefont{S.}~\bibnamefont{Bauer}},
  \bibinfo{journal}{Org. Electron.} \textbf{\bibinfo{volume}{7}},
  \bibinfo{pages}{188} (\bibinfo{year}{2006}).

\bibitem[{\citenamefont{Knupfer and Peisert}(2004)}]{E01}
\bibinfo{author}{\bibfnamefont{M.}~\bibnamefont{Knupfer}} \bibnamefont{and}
  \bibinfo{author}{\bibfnamefont{H.}~\bibnamefont{Peisert}},
  \bibinfo{journal}{phys. stat. sol. (a)} \textbf{\bibinfo{volume}{201}},
  \bibinfo{pages}{1055} (\bibinfo{year}{2004}).

\bibitem[{\citenamefont{Molodtsova et~al.}(2005)\citenamefont{Molodtsova,
  Schwieger, and Knupfer}}]{E02}
\bibinfo{author}{\bibfnamefont{O.~V.} \bibnamefont{Molodtsova}},
  \bibinfo{author}{\bibfnamefont{T.}~\bibnamefont{Schwieger}},
  \bibnamefont{and} \bibinfo{author}{\bibfnamefont{M.}~\bibnamefont{Knupfer}},
  \bibinfo{journal}{Appl. Surf. Sci.} \textbf{\bibinfo{volume}{252}},
  \bibinfo{pages}{143} (\bibinfo{year}{2005}).

\bibitem[{\citenamefont{Veenstra and Jonkman}(2003)}]{E03}
\bibinfo{author}{\bibfnamefont{S.~C.} \bibnamefont{Veenstra}} \bibnamefont{and}
  \bibinfo{author}{\bibfnamefont{H.~T.} \bibnamefont{Jonkman}},
  \bibinfo{journal}{J. Poly. Sci. B} \textbf{\bibinfo{volume}{41}},
  \bibinfo{pages}{2549} (\bibinfo{year}{2003}).

\bibitem[{\citenamefont{Molodtsova and Knupfer}(2006)}]{E04}
\bibinfo{author}{\bibfnamefont{O.~V.} \bibnamefont{Molodtsova}}
  \bibnamefont{and} \bibinfo{author}{\bibfnamefont{M.}~\bibnamefont{Knupfer}},
  \bibinfo{journal}{J. Appl. Phys.} \textbf{\bibinfo{volume}{99}},
  \bibinfo{pages}{053704} (\bibinfo{year}{2006}).

\bibitem[{\citenamefont{Hill et~al.}(2000)\citenamefont{Hill, Kahn, Soos, and
  Pascal}}]{GapCuPc}
\bibinfo{author}{\bibfnamefont{I.}~\bibnamefont{Hill}},
  \bibinfo{author}{\bibfnamefont{A.}~\bibnamefont{Kahn}},
  \bibinfo{author}{\bibfnamefont{Z.}~\bibnamefont{Soos}}, \bibnamefont{and}
  \bibinfo{author}{\bibfnamefont{R.}~\bibnamefont{Pascal}},
  \bibinfo{journal}{Chem. Phys. Lett.} \textbf{\bibinfo{volume}{327}},
  \bibinfo{pages}{181} (\bibinfo{year}{2000}).

\bibitem[{\citenamefont{Lof et~al.}(1992)\citenamefont{Lof, Vanveenendaal,
  Koopmans, Jonkman, and Sawatzky}}]{GapC60}
\bibinfo{author}{\bibfnamefont{R.~W.} \bibnamefont{Lof}},
  \bibinfo{author}{\bibfnamefont{M.~A.} \bibnamefont{Vanveenendaal}},
  \bibinfo{author}{\bibfnamefont{B.}~\bibnamefont{Koopmans}},
  \bibinfo{author}{\bibfnamefont{H.~T.} \bibnamefont{Jonkman}},
  \bibnamefont{and} \bibinfo{author}{\bibfnamefont{G.~A.}
  \bibnamefont{Sawatzky}}, \bibinfo{journal}{Phys. Rev. Lett.}
  \textbf{\bibinfo{volume}{68}}, \bibinfo{pages}{3924} (\bibinfo{year}{1992}).

\bibitem[{\citenamefont{Rand et~al.}(2005)\citenamefont{Rand, Xue, Uchida, and
  Forrest}}]{XRD_Mix}
\bibinfo{author}{\bibfnamefont{B.~P.} \bibnamefont{Rand}},
  \bibinfo{author}{\bibfnamefont{J.~G.} \bibnamefont{Xue}},
  \bibinfo{author}{\bibfnamefont{S.}~\bibnamefont{Uchida}}, \bibnamefont{and}
  \bibinfo{author}{\bibfnamefont{S.~R.} \bibnamefont{Forrest}},
  \bibinfo{journal}{J. Appl. Phys.} \textbf{\bibinfo{volume}{98}},
  \bibinfo{pages}{124902} (\bibinfo{year}{2005}).

\bibitem[{\citenamefont{Faiman et~al.}(1997)\citenamefont{Faiman, Goren, Katz,
  Koltun, Melnik, Shames, and Shtutina}}]{XRD_C60}
\bibinfo{author}{\bibfnamefont{D.}~\bibnamefont{Faiman}},
  \bibinfo{author}{\bibfnamefont{S.}~\bibnamefont{Goren}},
  \bibinfo{author}{\bibfnamefont{E.~A.} \bibnamefont{Katz}},
  \bibinfo{author}{\bibfnamefont{M.}~\bibnamefont{Koltun}},
  \bibinfo{author}{\bibfnamefont{N.}~\bibnamefont{Melnik}},
  \bibinfo{author}{\bibfnamefont{A.}~\bibnamefont{Shames}}, \bibnamefont{and}
  \bibinfo{author}{\bibfnamefont{S.}~\bibnamefont{Shtutina}},
  \bibinfo{journal}{Thin Solid Films} \textbf{\bibinfo{volume}{295}},
  \bibinfo{pages}{283} (\bibinfo{year}{1997}).

\bibitem[{\citenamefont{St{\"o}hr et~al.}(2001)\citenamefont{St{\"o}hr, Wagner,
  Gabriel, Weyers, and M{\"o}ller}}]{demixing}
\bibinfo{author}{\bibfnamefont{M.}~\bibnamefont{St{\"o}hr}},
  \bibinfo{author}{\bibfnamefont{T.}~\bibnamefont{Wagner}},
  \bibinfo{author}{\bibfnamefont{M.}~\bibnamefont{Gabriel}},
  \bibinfo{author}{\bibfnamefont{B.}~\bibnamefont{Weyers}}, \bibnamefont{and}
  \bibinfo{author}{\bibfnamefont{R.}~\bibnamefont{M{\"o}ller}},
  \bibinfo{journal}{Adv. Func. Mater.} \textbf{\bibinfo{volume}{11}},
  \bibinfo{pages}{175} (\bibinfo{year}{2001}).

\bibitem[{\citenamefont{Tuladhar et~al.}(2005)\citenamefont{Tuladhar,
  Poplavskyy, Choulis, Durrant, Bradley, and Nelson}}]{Tuladhar05}
\bibinfo{author}{\bibfnamefont{S.~M.} \bibnamefont{Tuladhar}},
  \bibinfo{author}{\bibfnamefont{D.}~\bibnamefont{Poplavskyy}},
  \bibinfo{author}{\bibfnamefont{S.~A.} \bibnamefont{Choulis}},
  \bibinfo{author}{\bibfnamefont{J.~R.} \bibnamefont{Durrant}},
  \bibinfo{author}{\bibfnamefont{D.~D.~C.} \bibnamefont{Bradley}},
  \bibnamefont{and} \bibinfo{author}{\bibfnamefont{J.}~\bibnamefont{Nelson}},
  \bibinfo{journal}{Adv. Func. Mater.} \textbf{\bibinfo{volume}{15}},
  \bibinfo{pages}{1171} (\bibinfo{year}{2005}).

\bibitem[{\citenamefont{Mihailetchi et~al.}(2005)\citenamefont{Mihailetchi,
  Koster, Blom, Melzer, de~Boer, van Duren, and Janssen}}]{Mihailetchi05}
\bibinfo{author}{\bibfnamefont{V.~D.} \bibnamefont{Mihailetchi}},
  \bibinfo{author}{\bibfnamefont{L.~J.~A.} \bibnamefont{Koster}},
  \bibinfo{author}{\bibfnamefont{P.~W.~M.} \bibnamefont{Blom}},
  \bibinfo{author}{\bibfnamefont{C.}~\bibnamefont{Melzer}},
  \bibinfo{author}{\bibfnamefont{B.}~\bibnamefont{de~Boer}},
  \bibinfo{author}{\bibfnamefont{J.~K.~J.} \bibnamefont{van Duren}},
  \bibnamefont{and} \bibinfo{author}{\bibfnamefont{R.~A.~J.}
  \bibnamefont{Janssen}}, \bibinfo{journal}{Adv. Func. Mater.}
  \textbf{\bibinfo{volume}{15}}, \bibinfo{pages}{795} (\bibinfo{year}{2005}).

\bibitem[{\citenamefont{Dinelli et~al.}(2006)\citenamefont{Dinelli, Capelli,
  Loi, Murgia, Muccini, Facchetti, and Marks}}]{Dinelli06}
\bibinfo{author}{\bibfnamefont{F.}~\bibnamefont{Dinelli}},
  \bibinfo{author}{\bibfnamefont{R.}~\bibnamefont{Capelli}},
  \bibinfo{author}{\bibfnamefont{M.~A.} \bibnamefont{Loi}},
  \bibinfo{author}{\bibfnamefont{M.}~\bibnamefont{Murgia}},
  \bibinfo{author}{\bibfnamefont{M.}~\bibnamefont{Muccini}},
  \bibinfo{author}{\bibfnamefont{A.}~\bibnamefont{Facchetti}},
  \bibnamefont{and} \bibinfo{author}{\bibfnamefont{T.~J.} \bibnamefont{Marks}},
  \bibinfo{journal}{Adv. Mater.} \textbf{\bibinfo{volume}{18}},
  \bibinfo{pages}{1416} (\bibinfo{year}{2006}).

\bibitem[{\citenamefont{Dinelli et~al.}(2004)\citenamefont{Dinelli, Murgia,
  Levy, Cavallini, Biscarini, and de~Leeuw}}]{Dinelli04}
\bibinfo{author}{\bibfnamefont{F.}~\bibnamefont{Dinelli}},
  \bibinfo{author}{\bibfnamefont{M.}~\bibnamefont{Murgia}},
  \bibinfo{author}{\bibfnamefont{P.}~\bibnamefont{Levy}},
  \bibinfo{author}{\bibfnamefont{M.}~\bibnamefont{Cavallini}},
  \bibinfo{author}{\bibfnamefont{F.}~\bibnamefont{Biscarini}},
  \bibnamefont{and} \bibinfo{author}{\bibfnamefont{D.~M.}
  \bibnamefont{de~Leeuw}}, \bibinfo{journal}{Phys. Rev. Lett.}
  \textbf{\bibinfo{volume}{92}}, \bibinfo{pages}{116802}
  (\bibinfo{year}{2004}).

\bibitem[{\citenamefont{Scheinert and Paasch}(2004)}]{Delta_VT}
\bibinfo{author}{\bibfnamefont{S.}~\bibnamefont{Scheinert}} \bibnamefont{and}
  \bibinfo{author}{\bibfnamefont{G.}~\bibnamefont{Paasch}},
  \bibinfo{journal}{phys. stat. sol. (a)} \textbf{\bibinfo{volume}{201}},
  \bibinfo{pages}{1263} (\bibinfo{year}{2004}).

\bibitem[{\citenamefont{Ruani et~al.}(2002)\citenamefont{Ruani, Fontanini,
  Murgia, and Taliani}}]{Ruani02}
\bibinfo{author}{\bibfnamefont{G.}~\bibnamefont{Ruani}},
  \bibinfo{author}{\bibfnamefont{C.}~\bibnamefont{Fontanini}},
  \bibinfo{author}{\bibfnamefont{M.}~\bibnamefont{Murgia}}, \bibnamefont{and}
  \bibinfo{author}{\bibfnamefont{C.}~\bibnamefont{Taliani}},
  \bibinfo{journal}{J. Chem. Phys.} \textbf{\bibinfo{volume}{116}},
  \bibinfo{pages}{1713} (\bibinfo{year}{2002}).

\bibitem[{\citenamefont{Chung and Neudeck}(1987)}]{TheoNeu1}
\bibinfo{author}{\bibfnamefont{K.~Y.} \bibnamefont{Chung}} \bibnamefont{and}
  \bibinfo{author}{\bibfnamefont{G.~W.} \bibnamefont{Neudeck}},
  \bibinfo{journal}{J. Appl. Phys.} \textbf{\bibinfo{volume}{62}},
  \bibinfo{pages}{4617} (\bibinfo{year}{1987}).

\bibitem[{\citenamefont{Neudeck et~al.}(1987)\citenamefont{Neudeck, Chung, and
  Bare}}]{TheoNeu2}
\bibinfo{author}{\bibfnamefont{G.~W.} \bibnamefont{Neudeck}},
  \bibinfo{author}{\bibfnamefont{K.~Y.} \bibnamefont{Chung}}, \bibnamefont{and}
  \bibinfo{author}{\bibfnamefont{H.~F.} \bibnamefont{Bare}},
  \bibinfo{journal}{IEEE Trans. Electron. Dev.} \textbf{\bibinfo{volume}{34}},
  \bibinfo{pages}{866} (\bibinfo{year}{1987}).

\bibitem[{\citenamefont{Neudeck and Malhotra}(1975)}]{a-Si-H}
\bibinfo{author}{\bibfnamefont{G.~W.} \bibnamefont{Neudeck}} \bibnamefont{and}
  \bibinfo{author}{\bibfnamefont{A.~K.} \bibnamefont{Malhotra}},
  \bibinfo{journal}{J. Appl. Phys.} \textbf{\bibinfo{volume}{46}},
  \bibinfo{pages}{239} (\bibinfo{year}{1975}).

\bibitem[{\citenamefont{Paasch et~al.}(2005)\citenamefont{Paasch, Lindner,
  Rost-Bietsch, Karg, Riess, and Scheinert}}]{TheoGP}
\bibinfo{author}{\bibfnamefont{G.}~\bibnamefont{Paasch}},
  \bibinfo{author}{\bibfnamefont{T.}~\bibnamefont{Lindner}},
  \bibinfo{author}{\bibfnamefont{C.}~\bibnamefont{Rost-Bietsch}},
  \bibinfo{author}{\bibfnamefont{S.}~\bibnamefont{Karg}},
  \bibinfo{author}{\bibfnamefont{W.}~\bibnamefont{Riess}}, \bibnamefont{and}
  \bibinfo{author}{\bibfnamefont{S.}~\bibnamefont{Scheinert}},
  \bibinfo{journal}{J. Appl. Phys.} \textbf{\bibinfo{volume}{98}},
  \bibinfo{pages}{084505} (\bibinfo{year}{2005}).

\bibitem[{\citenamefont{Smits et~al.}(2006)\citenamefont{Smits, Anthopoulos,
  Setayesh, van Veenendaal, Coehoorn, Blom, de~Boer, and de~Leeuw}}]{Smits06}
\bibinfo{author}{\bibfnamefont{E.~C.~P.} \bibnamefont{Smits}},
  \bibinfo{author}{\bibfnamefont{T.~D.} \bibnamefont{Anthopoulos}},
  \bibinfo{author}{\bibfnamefont{S.}~\bibnamefont{Setayesh}},
  \bibinfo{author}{\bibfnamefont{E.}~\bibnamefont{van Veenendaal}},
  \bibinfo{author}{\bibfnamefont{R.}~\bibnamefont{Coehoorn}},
  \bibinfo{author}{\bibfnamefont{P.~W.~M.} \bibnamefont{Blom}},
  \bibinfo{author}{\bibfnamefont{B.}~\bibnamefont{de~Boer}}, \bibnamefont{and}
  \bibinfo{author}{\bibfnamefont{D.~M.} \bibnamefont{de~Leeuw}},
  \bibinfo{journal}{Phys. Rev. B} \textbf{\bibinfo{volume}{73}},
  \bibinfo{pages}{205316} (\bibinfo{year}{2006}).

\end{thebibliography}
\end{document}